\documentclass[a4,10pt]{article}

\usepackage{amssymb}
\usepackage{graphicx}

\title{Flexible composition and execution of high performance, high 
       fidelity multiscale biomedical simulations}

\author{D. Groen$^{1}$, J. Borgdorff$^{2}$, C. Bona-Casas$^{2}$, J. Hetherington$^{1}$,\\ 
R.W. Nash$^{1}$, S.J. Zasada$^{1}$, I. Saverchenko$^{3}$, M. Mamonski$^{4}$,\\
K. Kurowski$^{4}$, M.O. Bernabeu$^{1}$, A.G. Hoekstra$^{2}$ \& P.V. Coveney$^{1}$\\
\small{$^{1}$ Centre for Computational Science, University College London, UK.}\\
\small{$^{2}$ Section Computational Science, University of Amsterdam, the Netherlands}\\
\small{$^{3}$ Leibniz-Rechenzentrum, Garching, Germany}\\
\small{$^{4}$ Poznan Supercomputing and Networking Center, Poznan, Poland}}

\date{}

\begin{document}

\maketitle

\begin{abstract} 
Multiscale simulations are essential in the biomedical domain
to accurately model human physiology. We present a modular
approach for designing, constructing and executing multiscale simulations on a
wide range of resources, from desktops to petascale supercomputers, including
combinations of these.  Our work features two multiscale applications, in-stent
restenosis and cerebrovascular bloodflow, which combine multiple existing
single-scale applications to create a multiscale simulation. These applications
can be efficiently coupled, deployed and executed on computers up to the largest
(peta) scale, incurring a coupling overhead of 1 to 10\% of the total
execution time.  
\end{abstract}

\section{Introduction}

Models of biomedical systems are inherently complex; properties on small time
and length scales, such as the molecular or genome level, can make a
substantial difference to the properties observed on much larger scales, such
as the organ, full-body and even the population level; and vice-versa~\cite{Sloot:2010,Noble:2009}. We
therefore need to apply multiscale approaches when modelling many biomedical
problems. Example biomedical multiscale challenges include predicting the
impact of a surgical procedure~\cite{Tahir:2011}, investigating the effects of pathologies (e.g.
arterial malformations or fistulas~\cite{Migliavacca:2005}), or assessing the effects of a
targeted drug on a given patient~\cite{ObiolPardo:2011}. In all these cases, we need to
examine processes that not only occur across several time and/or length scales, but
that also rely on different underlying physical and/or biological mechanisms. As a
result, modelling these processes may require substantially different algorithms 
and varying levels of computational effort. 

Historically, these problems have often been modelled using single scale
approaches, focussing exclusively on those aspects of the problem which are deemed most
relevant.  However, applying a single scale model is frequently insufficient 
to fully understand the problem at hand, as additional
processes occurring on different scales must be incorporated to obtain
sufficient accuracy. It is this need for understanding the composite
problem, rather than its individual subcomponents alone, that has driven many
research groups to explore multiscale
modelling; for example~\cite{Noble:2002,Finkelstein:2004,Hetherington:2007,Southern:2008}.

In a multiscale model, the overall system is approximated by coupling two or
more single scale {\em submodels}. Establishing and performing the data
exchange between these submodels is an important aspect of enabling multiscale
modelling.  It is often addressed by using coupling tools, such as
MUSCLE~\cite{Hegewald:2008,Borgdorff:2012}, the Multilevel Communicating
Interface~\cite{Grinberg:2011} and GridSpace~\cite{Ciepiela:2010}. Groen et
al.~\cite{Groen:2012-3} provide a review of coupling tools and the
computational challenges they address.

Another important aspect  is adopting a data standard which submodels can adopt
to exchange meaningful quantities. Several markup languages, such as
SBML~\cite{Hucka:2003} and CellML~\cite{Lloyd:2004}, resolve this problem by
providing a description language for the storage and exchange of model data and
submodel definitions.  CellML specifically allows for submodel exchange between
ODE and PDE solvers, whereas SBML is aimed towards biochemical pathway and
reaction ODE submodels. Both SBML and CellML, being languages for the
description of submodels and system data, have serious limitations in that they
require additional tools to perform tasks that are not directly related to
ensuring data interoperability. These include transferring data between submodels,
deploying submodels on appropriate resources and orchestrating the interplay of
submodels in a multiscale simulation. Additionally, they only provide very
limited features to describe submodels that do not rely on ODE-based methods,
such as finite-element/volume methods, spectral methods, lattice-Boltzmann,
molecular dynamics and particle-based methods, which are of increasing 
importance in the biomedical domain.

Here we present a comprehensive approach to enable multiscale biomedical
modelling from problem definition through the bridging of length and time scales,
to the deployment and execution of these models as multiscale
simulations. Our approach, which has been developed within the MAPPER
project~\footnote{http://www.mapper-project.eu/}, relies on coupling existing submodels and supports
the use of resources ranging from a local laptop to large international
supercomputing infrastructures, and distributed combinations of these.  We
present our approach for describing multiscale biomedical models in
section~\ref{Sec:multmodel} and for constructing and executing multiscale
simulations on large computing infrastructures in section~\ref{Sec:MAPPER}. We
describe our approach for biomedical applications in
section~\ref{Sec:reuse}. We have applied our approach to two biomedical
multiscale applications, in-stent restenosis and hierarchical cerebrovascular
blood flow, which we present in sections~\ref{Sec:ISR3D} and \ref{Sec:HemeLB}
respectively. We conclude with a brief discussion.

\section{Multiscale biomedical modelling}\label{Sec:multmodel}

Multiscale modelling gives rise to a number of challenges which extend beyond
the translation of model and system data. Most importantly, we seek to allow
application developers, such as computational biologists and biomedics, to couple multiscale
models for large problems, supporting any type of coupling, using any type of
submodel they wish to include, and executing this using any type of
computational resource, from laptop to petascale. Since we cannot expect
computational biologists to have expertise in all the technical details of
multiscale computing, we also aim to present a uniform and easy-to-use
interface which retains the power of the underlying technology. This also
enables users with less technical expertise to execute previously constructed
multiscale simulations.

Multiscale systems are, in general, characterized by the interaction of
phenomena on different scales, but the details vary for different
scientific domains. To preserve the generality of our approach, we adopt the
Multiscale Modelling and Simulation Framework (MMSF) to reason about multiscale
models in a domain-independent context and to create {\em recipes} for
constructing multiscale simulations independent of the underlying
implementations or computer architectures.  This MMSF is based on earlier work
on coupled cellular automata and agent-based
systems~\cite{Hoekstra:2007-2,Hoekstra:2010} and has been applied to several
computational problems in biomedicine~\cite{Evans:2008,Caiazzo:2011,Tahir:2011}.
Within the MMSF, the interactions between single-scale submodels are confined
to well-defined couplings. The submodels can therefore be studied as
independent models with dependent incoming and outgoing links. The graph of all
submodels and couplings, the coupling topology, can either be cyclic or
acyclic~\cite{Borgdorff:2011}.  In a cyclic coupling topology the submodels
will exchange information in an iterative loop, whereas in an
acyclic topology the submodels are activated one after another, resulting in a
directional data flow which makes them well-suited for workflow managers. Two
parts of the MMSF are particularly useful for our purposes, namely the {\em
Scale Separation Map}~\cite{Hoekstra:2007-2} and the {\em Multiscale Modelling
Language} (MML)~\cite{Falcone:2010,Borgdorff:2012}. The Scale Separation Map is
a graphical representation which provides direct insights into the coupling
characteristics of the multiscale application. MML provides a formalization of
the coupling between the submodels, independent of the underlying
implementation. In addition, it simplifies the issues associated with
orchestrating submodels by capturing the orchestration mechanisms in a simple
model which consists of only four distinct operators. MML definitions can be
stored for later use using an XML-based file format (xMML) or represented
visually using graphical MML (gMML)~\cite{Borgdorff:2012}.


The generic MML definitions allow us to identify commonalities between
multiscale applications in different scientific domains. Additionally, the
stored xMML can be used as input for a range of supporting tools that facilitate
multiscale simulations (e.g., tools that automate the deployment or
the submodel coupling of these simulations, as discussed in
section~\ref{Sec:MAPPER}). 

\section{From multiscale model to production simulation}\label{Sec:MAPPER}

We have developed a range of tools that allow us to create, deploy and run
multiscale simulations based on the xMML specification of a multiscale model.
Two of the main challenges in realising successful multiscale simulations
are to establish a fast and flexible coupling between submodels and to deploy
and execute the implementations of these submodels efficiently on computer
resources of any size. In this section we review these challenges and present
our solutions to them. 

\subsection{Coupling submodels}

Effective and efficient coupling between submodels encompasses three
major aspects. First, we must translate our MML definitions to technical recipes
for the multiscale simulation execution.  Second, we need to
initiate and orchestrate the execution of the submodels in an automated way,
in accordance with the MML specification. Third, we need to 
efficiently exchange model and system information between the submodels. 

The Multiscale Library and Coupling Environment
(MUSCLE)~\cite{Hegewald:2008,Borgdorff:2012} provides a solution to the first
two aspects. It uses MML in conjunction with a definition of the requested
resources to bootstrap and start the different submodels and to establish a
connection for data exchange between the submodels. It also implements the
coupling definitions defined in the MMSF to orchestrate the submodels. The
submodels can be coupled either locally (on a workstation for example), via a local
network, or via a wide area network using the MUSCLE Transport Overlay
(MTO)~\cite{Zasada:2012}. Running submodels in different locations and coupling
them over a wide area network is especially important when submodels have
very different computational requirements, for example when one submodel
requires a local machine with a set of Python libraries, and another submodel
requires a large supercomputer. However, messages exchanged across a wide 
area network do take longer to arrive. Among other things, MTO provides the means to
exchange data between large supercomputers while adhering to the local security
policies and access restrictions.

The exchanges between submodels overlap with computations in many cases
(see section~\ref{Sec:HemeLB} for an example), allowing for an efficient
execution of the overall simulation. However, when the exchanges are
particularly frequent or the exchanged data particularly large, inefficiencies
in the data exchange can severely slow down the multiscale simulation as a
whole. We use the MPWide communication library~\cite{mpwide}, previously used
to run cosmological simulations distributed across supercomputers~\cite{sushi},
to optimise our wide area communications for performance. We already use MPWide
directly within the cerebrovascular bloodflow simulation and are currently
incorporating it in MUSCLE to optimise the performance of the MTO.

\subsection{Deploying and executing submodels on production resources}

When large multiscale models are deployed and executed as simulations on
production compute resources, a few major practical challenges arise. 
Production resources are often shared by many users, and are difficult to use
even partially at appropriate times. This in turn makes it difficult to ensure
that the submodels are invoked at the right times, and in cases where they are
distributed on different machines, to ensure that the multiscale simulation
retains an acceptable time to completion. We use the QosCosGrid environment
(QCG)~\cite{Kurowski:2010} to run submodels at a predetermined time on large
computing resources. QCG enables us, if installed on the target resource, to
reserve these resources in advance for a given period of time to allow our
submodels to be executed there. This is valuable in applications that require
cyclic coupling, as we then require multiple submodels to be executed either
concurrently or in alternating fashion. Additionally, with QCG we can
explicitly specifiy the sequence of resources to be reserved, allowing us to
repeat multiple simulations in a consistent manner using the same resource
sequence each time. A second component that aids in the execution of submodels
is the Application Hosting Environment (AHE)~\cite{Zasada:2009}. AHE simplifies
user access to production resources by streamlining the authentication methods
and by centralising the installation and maintenance tasks of application
codes. The end users of AHE only need to work with one uniform client interface to
access and run their applications on a wide range of production resources.

\section{Using MAPPER for biomedical problems}\label{Sec:reuse}

We apply the MAPPER approach to a number of applications. We present
two of these applications here (in-stent restenosis and cerebrovascular blood
flow), but we have also used MAPPER to define, deploy and execute multiscale
simulations of clay-polymer nanocomposite materials~\cite{Suter:2012}, river beds and
canals~\cite{BenBelgacem:2012}, and several problems in nuclear fusion.

MAPPER provides formalisms, tools and services which aid in the description of
multiscale models, as well as the construction, deployment and execution of
multiscale simulations on production infrastructures. It is intended as a
general-purpose solution, and as such tackles challenges in multiscale
simulation that exist across scientific disciplines. There are a number of
challenges which are outside the scope of our approach, because they may
require different solutions for different scientific disciplines. These include
choosing appropriate submodels for a multiscale simulation and defining, on
the application level, what information should be exchanged between submodels
at which times to provide a scientifically accurate and stable multiscale simulation.
However, MMSF does simplify the latter task by providing a limited
number of orchestration mechanisms.  The formalisms, tools and services
presented here are independent components, allowing users to adopt those parts
of our approach which  specifically meet their requirements in multiscale
modelling. This modular approach makes it easier for users to exploit the
functionalities of individual components and helps to retain a lightweight
simulation environment.

\section{Modelling in-stent restenosis}\label{Sec:ISR3D}

Coronary heart disease (CHD) is one of the most common causes of death, and 
is responsible for about 7.3 million deaths per year worldwide~\cite{WHO_CVD_report:2012}.  
CHD is typically expressed as artherosclerosis,
which corresponds with a thickening and hardening of blood vessels caused by
build-up of atheromatous plaque; when this significantly narrows the vessel, it
is called a stenosis. A common intervention for stenosis is stent-assisted
balloon angioplasty where a balloon, attached to a stent, is inserted in the
blood vessel and inflated at the stenosed location, consequently deploying the
stent. The stent acts as a scaffold for the blood vessel, compressing
the plaque and holding the lumen open. Occasionally, however, this intervention
is followed by in-stent restenosis (ISR), an excessive regrowth of tissue due
to the injury caused by the stent deployment~\cite{Moustapha:2001,Kastrati:2000}. 
Although there are a number of different
hypotheses~\cite{Jukema:2011}, the pathophysiological mechanisms and risk
factors of in-stent restenosis are not yet fully clear.

\begin{figure}
  \centering
  \includegraphics[width=0.45\textwidth]{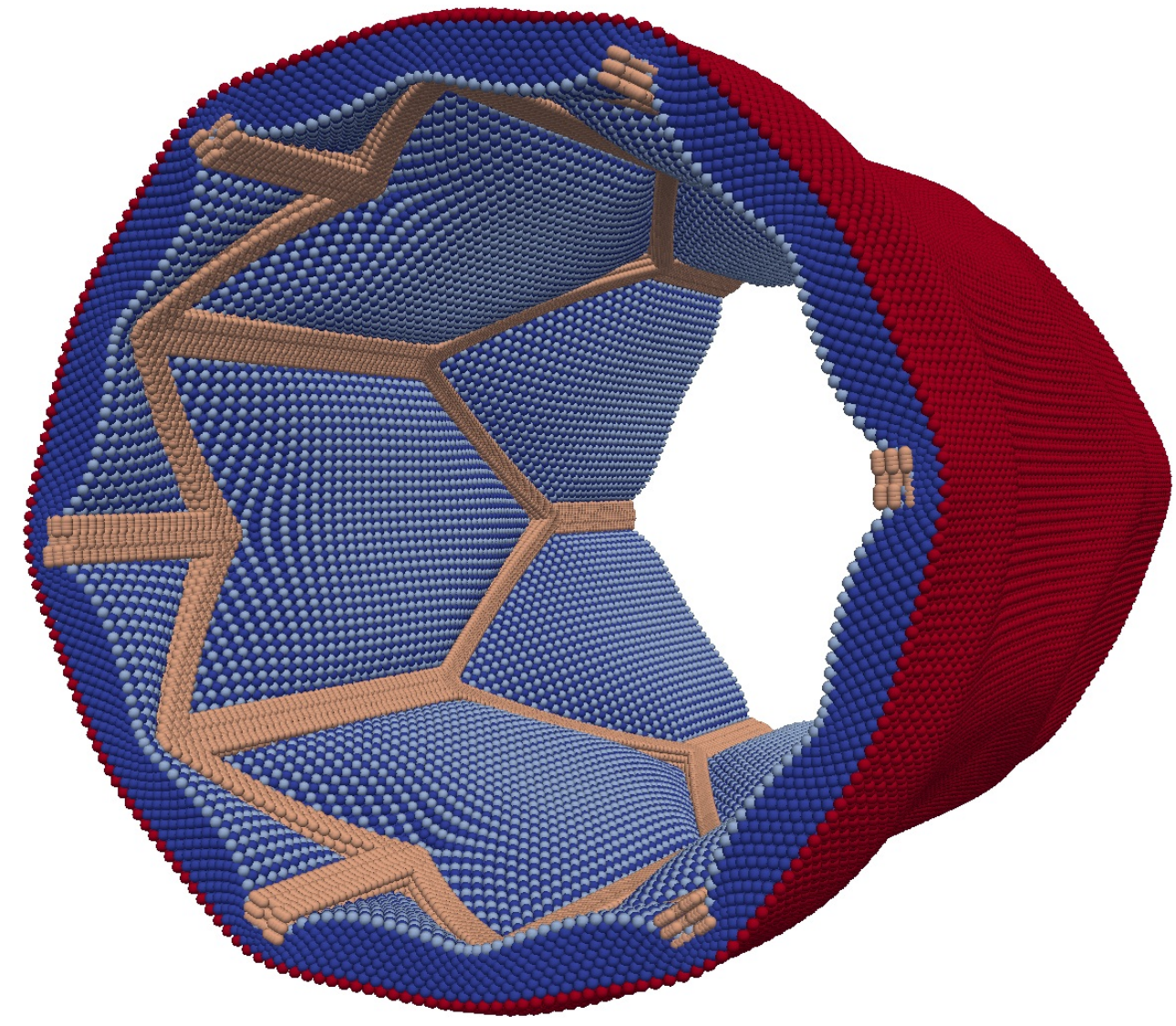}
  \includegraphics[width=0.45\textwidth]{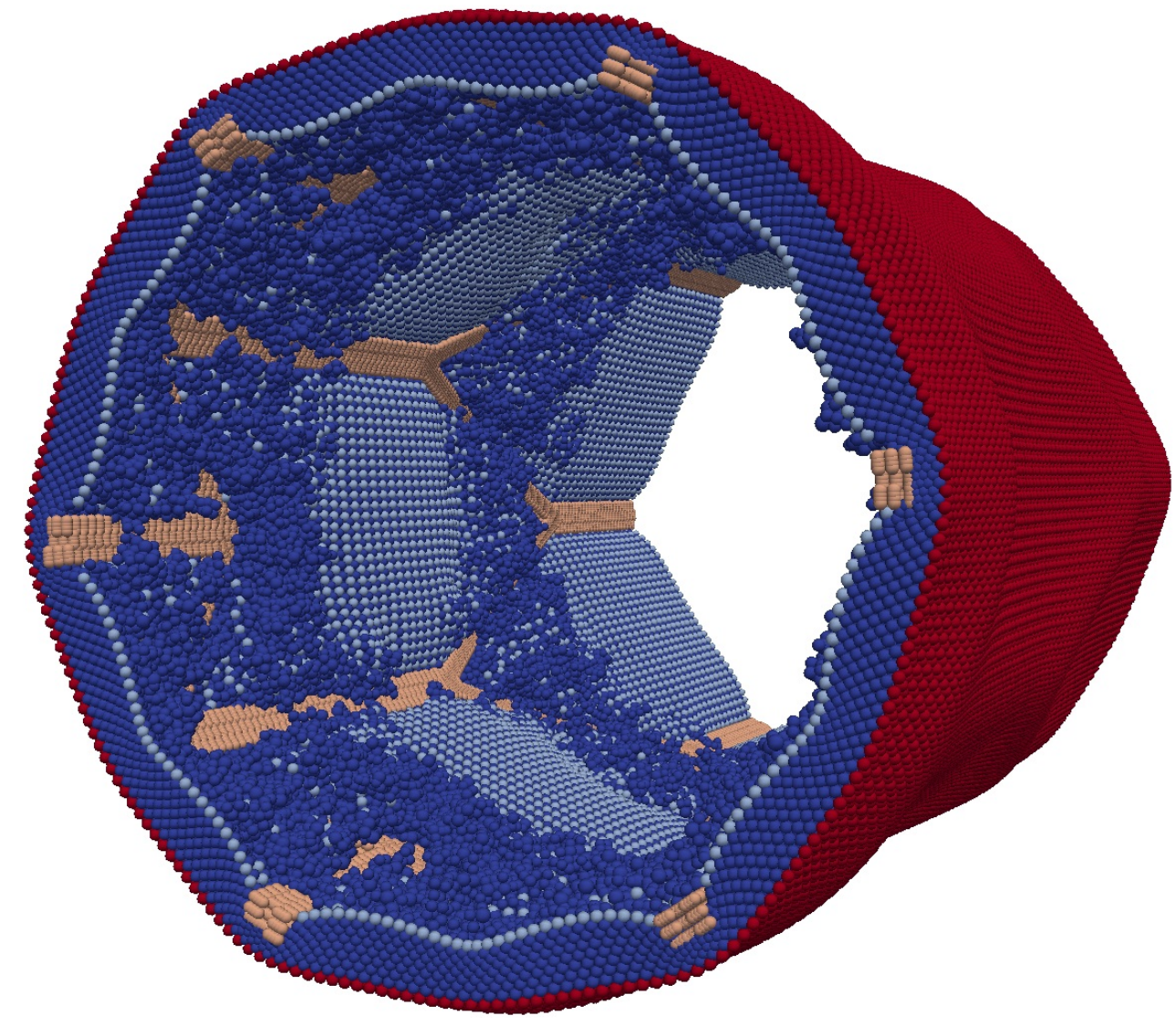}
  \caption{Images of a stented artery as modelled using ISR3D, before
restenosis occurs on the left and after 12.5 days of muscle cell proliferation 
on the right.}\label{fig:isr:smc}
\end{figure}

By modelling in-stent restenosis with a three-dimensional model (ISR3D) it is
possible to test mechanisms and risk factors that are likely to be the main
contributors to in-stent restenosis.  After evaluating the processes involved
in in-stent restenosis \cite{Evans:2008}, ISR3D applies the hypothesis that
smooth muscle cell proliferation drives the restenosis, and that this is
affected most heavily by wall shear stress of the blood flow, which regulates
endothelium recovery, and by growth inhibiting drugs diffused by a drug-eluting
stent. Using the model we can evaluate the effect of different drug
intensities, physical stent designs, vascular geometries and endothelium
recovery rates. ISR3D is derived from the two-dimensional
ISR2D~\cite{Caiazzo:2011,Tahir:2011} code. Compared to ISR2D it provides
additional accuracy, incorporating a full stent design, realistic cell growth,
and a three-dimensional blood flow model, although it requires more
computational effort. We present several results from modelling a stented
artery using ISR3D in figure~\ref{fig:isr:smc}. This figure shows the artery
both before and after restenosis has occurred.

\begin{figure}
  \centering
  \includegraphics[width=0.95\textwidth]{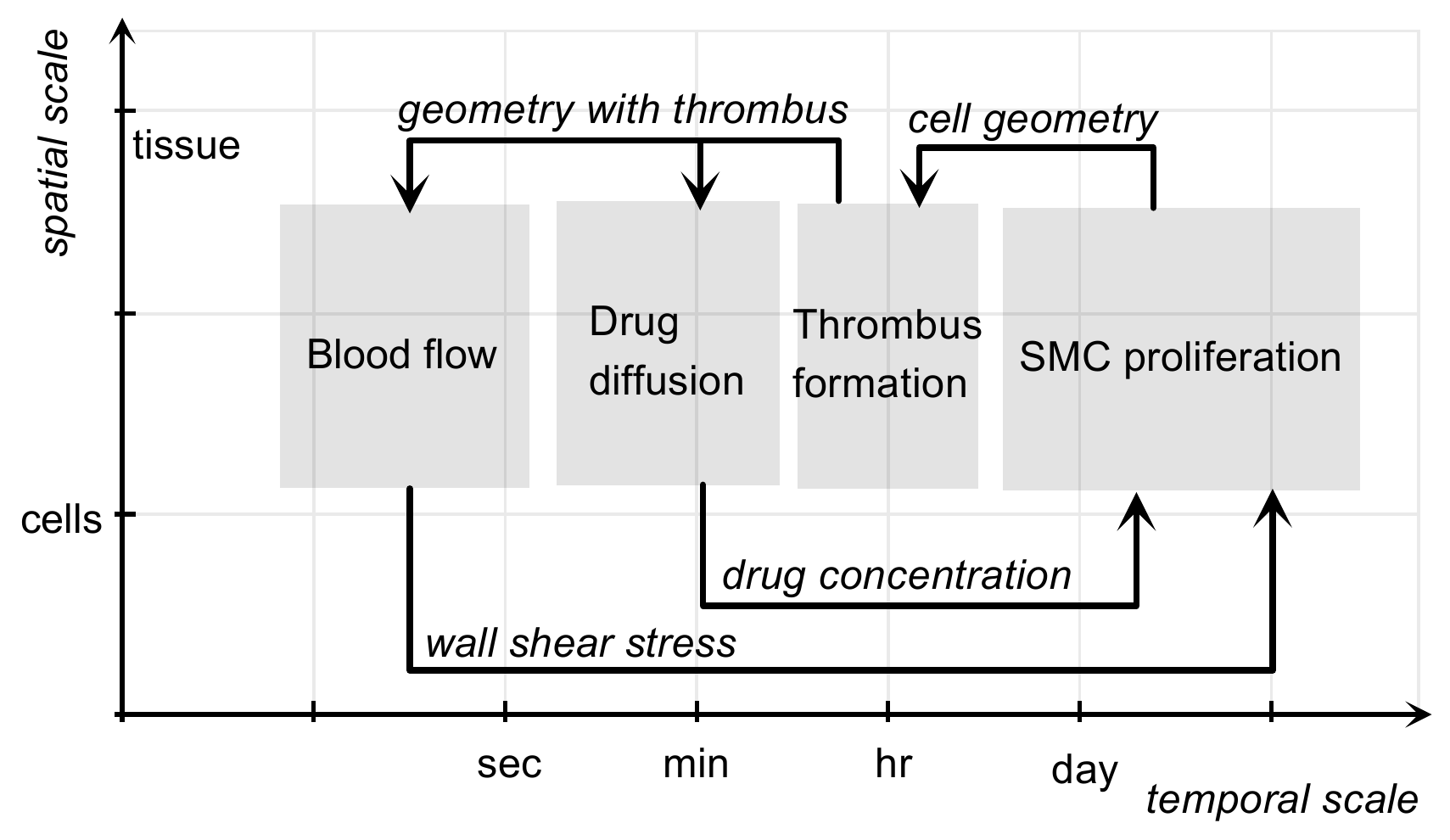}
  \caption{Scale Separation Map of the ISR3D multiscale simulation, originally 
  presented in~\cite{Borgdorff:2012}.}\label{fig:isr:ssm}
\end{figure}

From a multiscale modelling perspective, ISR3D combines four submodels, each
operating on a different time scale: smooth muscle cell proliferation (SMC),
which models the cell cycle, cell growth, and physical forces between cells;
initial thrombus formation (ITF) due to the backflow of blood; drug diffusion
(DD) of the drug eluding stent through the tissue and applied to the smooth
muscle cells; and blood flow (BF) and the resulting wall shear stress on smooth
muscle cells. We show the time scales of each submodel and the relations
between them in figure~\ref{fig:isr:ssm}.  The SMC submodel uses an agent-based
algorithm on the cellular scale, which undergoes validation on the tissue
level. All other submodels act on a cartesian grid representation of those
cells. For the BF submodel we use 1,237,040 lattice sites, and for the SMC
submodel we use 196,948 cells. The exchanges between the submodels are in the
order of 10 to 20 MB. The SMC sends a list of cell locations and sizes 
(stored as 8-byte integers) to the ITF, which sends the geometry (stored as
a 3D matrix of 8-byte integers) to BF and DD. In turn, BF and DD respectively send a list of wall shear
stress and drug concentrations (stored as 8-byte doubles) to SMC. Each 
coupling communication between SMC and  the other submodels takes place
once per SMC iteration. 

The submodels act independently, apart from exchanging messages, and are
heterogeneous. The SMC code is implemented in C++, the DD code in Java, and the
ITF code in Fortran. The BF code, which unlike the other codes runs in parallel,
uses the Palabos lattice-Boltzmann
application\footnote{http://www.palabos.org/}, written in C++.  The MML
specification of ISR3D contains a few conversion modules not mentioned above,
which perform basic data transformations necessary to ensure that the various
single scale models do not need to be aware of other submodels and their
internal representation or scales. Specifically, the submodels operate within 
the same domain, requiring the application to keep the grid and cell 
representation consistent, as well as their dimensions.

\subsection{Tests}

We have performed a number of tests to measure both the runtime and the
efficiency of our multiscale ISR3D simulation. The runs that were performed 
here were short versions of the actual simulations, with a limited number of 
iterations. The runs constributed to the integration testing of the code and 
allowed us to estimate the requirements for future computing resource proposals. 

We have run our tests in five different scenarios, using the EGI resource in
Krakow (Zeus, one scenario), a Dutch PRACE tier-1 machine in Amsterdam
(Huygens, two scenarios), or a combination of both (two scenarios). We provide
the technical specifications of the resources in table~\ref{table:machines}.
Since the runtime behaviour of ISR3D is cyclic, determined by the number of
smooth muscle cell iterations, we measured the runtime of a single cycle for
each scenario. Because only the BF model is parallelised, the resources we use
are partially idle whenever we are not executing the BF model. To reduce this
overhead of `idle time', we created two {\em double mapping} scenarios, each of
which runs two simulations in alternating fashion using the same resource
reservation. In these cases, the blood flow calculations of one simulation
takes place while the other simulation executes one or more of the other
submodels. We use a wait/notify signalling system within MUSCLE to enforce that
only one of the two simulations indeed executes its parallel BF model. We have
run one double mapping scenario locally on Huygens, and one distributed over
Huygens and Zeus. 

\begin{table}[hbt]
\centering
\begin{tabular}{llcccll}
name & processor & freq. & cores & mem/node & middleware\\
\hline
HECToR  & AMD Interlagos & 2.3 GHz & 512/2048 & 32 GB  & UNICORE\\
Huygens & IBM Power6     & 4.7 GHz & 32 & 128 GB & UNICORE\\
Henry   & Intel Xeon     & 2.4 GHz & 1  & 6 GB   & none\\ 
Zeus    & Intel Xeon     & 2.4 GHz & 4  & 16 GB  & QCG-Comp.\\
\end{tabular}
\caption{Computational characteristics of the machines we use in our multiscale simulations for
ISR3D and HemeLB. Clock frequency is given in the second column, the number of cores used in the
third column, and the amount of memory per node in the fourth column. The administrative details 
are listed in table~\ref{table:resources}.}\label{table:machines}
\end{table}

\begin{table}[hbt]
\centering
\begin{tabular}{llll}
name & provider & location & infrastructure\\
\hline
HECToR  & EPCC & Edinburgh, United Kingdom  & PRACE Tier-1\\
Huygens & SARA & Amsterdam, The Netherlands & PRACE Tier-1\\
Henry   & UCL  & London, United Kingdom     & Local workstation\\
Zeus    & Cyfronet & Krakow, Poland         & EGI (PL-Grid)\\
\end{tabular}
\caption{Administrative information for the resources described in table~\ref{table:machines}.}\label{table:resources}
\end{table}

\subsubsection{Results}

\begin{table}
    \centering
\begin{tabular}{lccccc}
scenario & BF & other & coupling & total & usage \%\\
\hline
Zeus                & 2100 & 1140 & 79  & 3240 & 80\% \\
Huygens             & 480  & 1260 & 73  & 1813 & 27\% \\
Huygens-double      & 480  & 1320 & 131 & 1931 & 45\% \\
Zeus-Huygens        & 480  & 960  & 92  & 1532 & 32\% \\
Zeus-Huygens-double & 480  & 1080 & 244 & 1804 & 56\% \\
\end{tabular}

    \caption{Runtimes with different scenarios. The name of the scenario is
given in the first column, the time spent on BF in the second, and the total
time spent on the other submodels in the third column (both rounded to the
nearest minute). We provide the coupling overhead in the fourth column and the
total simulation time per cycle in the fifth column. In the sixth column we
provide the average fraction of reserved resources doing computations (not idling)
throughout the period of execution. All time are given in seconds.}\label{table:isrruntime}

\end{table}

We present our performance measurements for the five scenarios in
table~\ref{table:isrruntime}. The runtime of our simulation is reduced by almost
50\% when we use the Huygens machine instead of Zeus. This is because we run the
BF code on Huygens using 32 cores, and on Zeus using 4 cores. However, the
usage of the reservation was considerably lower on Huygens than on Zeus for two
reasons: first, the allocation on Huygens was larger, leaving more cores idle
when the sequential submodels were computed; second, the ITF solver uses the
gfortran compiler, which is not well optimized for Huygens architecture. As a
result it requires only 4 minutes on Zeus and 12 minutes on Huygens. The
performance of the other sequential submodels is less platform dependent. When
we run our simulation distributed over both resources, we achieve the lowest
runtime. This is because we combine the fast execution of the BF module on
Huygens with the fast execution of the ITF solver on Zeus.  

When we use double mapping, the runtime per simulation increases by about 6\%
(using only Huygens) to 18\% (using both resources). However, the
double-mapping improves the usage of the reservation by a factor of $\sim1.7$.
Double-mapping is therefore an effective method to improve the resource usage
without incurring major increases to the time to completion of each simulation.
The coupling overhead is relatively low throughout our runs, consisting
of no more than 11\% of the runtime throughout our measurements. Using our 
approach, we are now modelling different in-stent restenosis scenarios, 
exploring a range of modelling parameters. 

\subsection{Clinical directions}

We primarily seek to understand which biological pathways dominate in
the process leading to in-stent restenosis. If successful, this will have two
effects on clinical practice: first, it suggests which factors are important for
in-stent restenosis, in turn giving clinicians more accurate estimates of
what the progression of the disease can be; second, it may spur further
directed clinical research of certain pathways, which will help the next
iteration of the model give more accurate results.
 
The methods for achieving this divide naturally in two directions: general
model validation and experiments; and virtual patient cohort studies. For
general model validation we consult the literature and use basic experimental
data, such as measurements from animal studies. In addition, we intend to use
virtual patient cohort studies to assess the in-stent restenosis risk factors
of virtual patients with different characteristics. In clinical practice,
this will not lead to personalized estimates, but rather to patient classifiers
on how ISR3D will progress. Once the primary factors leading to in-stent
restenosis have been assessed, a simplified model could be made based on ISR3D,
which takes less computational effort and runs within a hospital.

\section{Modelling of cerebrovascular blood flow}\label{Sec:HemeLB}

Our second hemodynamic example aims to incorporate not only the local arterial
structure in our models, but also properties of the circulation in the rest of
the human body. The key feature of this application is the multiscale modelling
of blood flow, delivering accuracy in the regions of direct scientific or
clinical interest while incorporating global blood flow properties using
more approximate methods. Here we provide an overview of our multiscale model
and report on its performance.  A considerable amount of previous work has been
done where groups combined bloodflow solvers of different types, for example in
the area of cardiovascular~\cite{Shi:2011,VanDeVosse:2011,Liu:2012} or
cerebrovascular bloodflow~\cite{Alastruey:2008,Grinberg:2011}.

We have constructed a distributed multiscale model in which we combine the open
source HemeLB lattice-Boltzmann application for blood flow modelling in
3D~\cite{Mazzeo:2008,Carver:2012} with the open source one-dimensional Python Navier-Stokes
(pyNS) blood flow solver~\cite{Botti:2010}. HemeLB is optimized for sparse
geometries such as vascular networks, and has been shown to scale linearly up
to at least 32,768 cores~\cite{Groen:2012}.  PyNS is a discontinuous Galerkin
solver which is geared towards modelling large arterial structures.  It uses
aortic blood flow input based on a set of patient-specific parameters and it
combines 1D wave propagation elements to model arterial vasculature with 0D
resistance elements to model veins. The numerical code supports thread-level
parallelization, and is written in Python in conjunction with the numpy
numerical library.

\subsection{Simulations}

We have run a number of coupled simulations using both HemeLB and pyNS as
submodels. Within pyNS we use a customised version of the 'Willis' model, based
on~\cite{Mulder:2011}, with a mean pressure 90 mmHg, a heart rate of 70 beats
per minute and a cardiac output of 5.68 litres per minute. Our model includes the
major arteries in the human torso, head and both arms, as well as a full model
for the circle of Willis, which is a major network of arteries in the human
head. For pyNS we use a time step size of $2.3766 \times 10^{-4}$ s.

We have modified a section of the right mid-cerebral artery (MCA) in pyNS to
allow it to be coupled to HemeLB in four places, exchanging pressure values in
these boundary regions. In HemeLB we simulate a small network of arteries, with
a voxel size of $3.5 \times 10^{-5}$ m and consisting of about 4.2 million
lattice sites which occupy 2.3\% of the simulation box volume. The HemeLB
simulation runs with a timestep of $2.3766 \times 10^{-6}$ s, a Mach number of
0.1 and a relaxation parameter $\tau$ of 0.52. We run pyNS using a local
machine at UCL (Henry), while we run HemeLB for 400,000 time steps on the
HECToR supercomputer in Edinburgh. The round-trip time for a network message
between these two resources is on average 11 milliseconds. We provide technical
details of both machines in table~\ref{table:resources}.  Both codes exchange pressure data at an
interval of 100 HemeLB time steps (or 1 pyNS time step). Because HemeLB time 
steps can take as little as 0.0002 seconds, we adopted MPWide to connect our submodels,
which run concurrently, and minimize the communication response time. The  exchanged 
data is represented using 8-byte doubles, and has a small
aggregate  size (less than 1 kb).  As a comparison, we have also run HemeLB as a standalone 
single scale simulation (labelled `ss'), retrieving its boundary values from a local configuration file. The pyNS code
requires 116 seconds to simulate 4000 time steps of our modified circle of
Willis problem when run as a stand-alone code.

\begin{table}
    \centering

\begin{tabular}{llrrrrrr}
ss/ms & cores        & Init & HemeLB* & coupling & total & coupling\\
      &              & t/s  & t/s     & t/s      & t/s   & efficiency\\
\hline
ss    & 512          & 47.9 & 2223    & n/a      & 2271  & n/a      \\
ms    & 512          & 51.2 & 2223    & 24       & 2298  & 98.8\%   \\
ss    & 2048         & 46.4 & 815     & n/a      & 862   & n/a      \\
ms    & 2048         & 50.5 & 815     & 37       & 907   & 95.0\%   \\
\end{tabular}
\caption{ Performance measurements of coupled simulations which consist of
HemeLB running on HECToR and pyNS running on a local UCL workstation. The type of
simulation, single scale (ss) or coupled multiscale (ms) is given in the first
colums. The number of cores used by HemeLB and the initialisation time are
respectively given in the second and third column. The time spent on HemeLB and
coupling work are given respectively in the fourth and fifth column, the total
time in the sixth column and the efficiency in the seventh column. The times
for HemeLB model execution in the multiscale runs are estimates, which we derived 
directly from the single scale performance results of the same problem on the 
same resources.
}\label{table:hemelbperf}

\end{table}

We present our results in table~\ref{table:hemelbperf}. For 512 cores, the
single scale HemeLB simulation takes 2271 seconds to perform its 400,000
lattice-Boltzmann time steps. The coupled HemeLB-pyNS simulation is only
marginally slower, reaching completion in 2298 seconds. We measure a coupling
overhead of only 24 seconds, which is the time to do 4,002 pressure exchanges
between HemeLB and pyNS. This amounts to about 6 milliseconds per exchange,
well below even the round-trip time of the network between UCL and EPCC alone.
The communication time on the HemeLB side is so low because pyNS runs faster
per coupling iteration than HemeLB, and the incoming pressure values are
already waiting at the network interface when HemeLB begins to send out its
own. As a result, the coupling overhead is only 24 seconds, and the multiscale
simulation is only 1.2\% slower than a single-scale simulation of the same
network domain.

When using 2048 cores, the runtime for the single-scale HemeLB simulation is
815 seconds, which is a speedup of 2.63 compared to the 512 core run. The
coupling overhead of the multiscale simulation is relatively higher than that
of the 512 core run due to a larger number of processes with which pressures
must be exchanged.  This results in an overall coupling efficiency of $0.95$,
which is lower than for the 512 core run. However, the 2048 core run contains
$\sim2000$ sites per core, which is a regime where we no longer achieve linear
scalability~\cite{Groen:2012} to begin with. As a future task, we plan to
coalesce these pressure exchanges with the other communications in HemeLB,
using the coalesced communication pattern~\cite{Carver:2012-2}.


\subsection{Clinical directions}

We aim to understand the flow dynamics in cerebrovascular networks and to predict
the flow dynamics in brain aneurysms for individual patients. The ability to 
predict the flow dynamics in cerebrovascular networks is of practical use to 
clinicians, as it allows them to more accurately determine whether surgery is 
required for a specific patient suffering from an aneurysm. In addition, our 
work supports a range of other scenarios, which in turn may drive clinical 
investigations of other vascular diseases.

Our current efforts focus on enhancing the model by introducing velocity
exchange and testing our models for accuracy. As a first accuracy test we
compared different boundary conditions and flow models within
HemeLB~\cite{Carver:2012}. Additionally we used our HemeLB-pyNS setup to
compare different blood rheology models, which we describe in detail
in~\cite{Bernabeu:2012}. Next steps include more patient-specific studies,
where we wish to compare the flow behavior within patient-specific networks of
arteries in our multi-scale simulations with the measured from those same
patients. We have already established several key functionalities to allow
patient-specific modelling. For example, HemeLB is able to convert 3D
rotational angiographic data into initial conditions for the simulation, while
pyNS provides support for patient-specific global parameters and customized
arterial tree definitions.  

\section{Conclusions and Future Work}\label{Sec:conc}

We have shown that MAPPER provides a usable and modular environment which enables 
us to efficiently map multiscale simulation models to a range of computing
infrastructures. Our methods provide computational biologists with the ability to
more clearly reason about multiscale simulations, to formally define their 
multiscale scenarios, and to more quickly simulate problems that involve the use 
of multiple codes using a range of resources. We have presented two applications,
in-stent restenosis and cerebrovascular bloodflow, and conclude that both 
applications run rapidly and efficiently, even when using multiple compute resources
in different geographical locations.

\section*{Acknowledgements}

We thank our colleagues in the MAPPER consortium, the HemeLB development team
at UCL, as well as Simone Manini and Luca Antiga from the pyNS development team.  This
work received funding from the MAPPER EU-FP7 project (grant no. RI-261507) and
the CRESTA EU-FP7 project (grant no. RI-287703).  This work made use of
computational resources provided by the PL-Grid Infrastructure (Zeus), by PRACE at 
SARA in Amsterdam, the Netherlands (Huygens) and EPCC in Edinburgh, United Kingdom
(HECToR), and by University College London (Henry).

\bibliographystyle{plain}
\bibliography{Library}

\begin{thebibliography}{10}

\bibitem{Sloot:2010}
P.~M.~A. Sloot and A.~G. Hoekstra.
\newblock Multi-scale modelling in computational biomedicine.
\newblock {\em Briefings in Bioinformatics}, 11(1):142--152, 2010.

\bibitem{Noble:2009}
P.~Kohl and D.~Noble.
\newblock {Systems biology and the virtual physiological human}.
\newblock {\em Molecular Systems Biology}, 5(1), July 2009.

\bibitem{Tahir:2011}
H.~Tahir, A.~G. Hoekstra, E.~Lorenz, P.~V. Lawford, D.~R. Hose, J.~Gunn, and
  D.~J.~W. Evans.
\newblock Multi-scale simulations of the dynamics of in-stent restenosis:
  impact of stent deployment and design.
\newblock {\em Interface Focus}, 1(3):365--373, 2011.

\bibitem{Migliavacca:2005}
F.~Migliavacca and G.~Dubini.
\newblock Computational modeling of vascular anastomoses.
\newblock {\em Biomechanics and Modeling in Mechanobiology}, 3:235--250, 2005.

\bibitem{ObiolPardo:2011}
C.~Obiol-Pardo, J.~Gomis-Tena, F.~Sanz, J.~Saiz, and M.~Pastor.
\newblock A multiscale simulation system for the prediction of drug-induced
  cardiotoxicity.
\newblock {\em Journal of Chemical Information and Modeling}, 51(2):483--492,
  2011.

\bibitem{Noble:2002}
D.~Noble.
\newblock Modeling the heart--from genes to cells to the whole organ.
\newblock {\em Science}, 295(5560):1678--1682, 2002.

\bibitem{Finkelstein:2004}
A.~Finkelstein, J.~Hetherington, L.~Li, O.~Margoninski, P.~Saffrey, R.~Seymour,
  and A.~Warner.
\newblock Computational challenges of systems biology.
\newblock {\em Computer}, 37:26--33, May 2004.

\bibitem{Hetherington:2007}
J.~Hetherington, I.~D.~L. Bogle, P.~Saffrey, O.~Margoninski, L.~Li,
  M.~Varela~Rey, S.~Yamaji, S.~Baigent, J.~Ashmore, K.~Page, R.~M. Seymour,
  A.~Finkelstein, and A.~Warner.
\newblock Addressing the challenges of multiscale model management in systems
  biology.
\newblock {\em Computers and Chemical Engineering}, 31(8):962 -- 979, 2007.

\bibitem{Southern:2008}
J.~Southern, J.~Pitt-Francis, J.~Whiteley, D.~Stokeley, H.~Kobashi, R.~Nobes,
  Y.~Kadooka, and D.~Gavaghan.
\newblock Multi-scale computational modelling in biology and physiology.
\newblock {\em Progress in Biophysics and Molecular Biology}, 96(1-3):60 -- 89,
  2008.

\bibitem{Hegewald:2008}
J.~Hegewald, M.~Krafczyk, J.~T\"{o}lke, A.~Hoekstra, and B.~Chopard.
\newblock An agent-based coupling platform for complex automata.
\newblock In {\em Proceedings of the 8th international conference on
  Computational Science, Part II}, ICCS '08, pages 227--233, Berlin,
  Heidelberg, 2008. Springer-Verlag.

\bibitem{Borgdorff:2012}
J.~Borgdorff, C.~Bona-Casas, M.~Mamonski, K.~Kurowski, T.~Piontek, B.~Bosak,
  K.~Rycerz, E.~Ciepiela, T.~Guba{\l}a, and D.~et~al. Harezlak.
\newblock {A Distributed Multiscale Computation of a Tightly Coupled Model
  Using the Multiscale Modeling Language}.
\newblock {\em Procedia Computer Science}, 9:596--605, 2012.

\bibitem{Grinberg:2011}
L.~Grinberg, V.~Morozov, D.~Fedosov, J.A. Insley, M.E. Papka, K.~Kumaran, and
  G.E. Karniadakis.
\newblock A new computational paradigm in multiscale simulations: Application
  to brain blood flow.
\newblock In {\em High Performance Computing, Networking, Storage and Analysis
  (SC), 2011 International Conference for}, pages 1--12, nov. 2011.

\bibitem{Ciepiela:2010}
E.~Ciepiela, D.~Harẹżlak, J.~Kocot, T.~Bartyński, M.~Kasztelnik,
  P.~Nowakowski, T.~Gubala, M.~Malawski, and M.~Bubak.
\newblock Exploratory programming in the virtual laboratory.
\newblock In {\em Computer Science and Information Technology (IMCSIT),
  Proceedings of the 2010 International Multiconference on}, pages 621 --628,
  oct. 2010.

\bibitem{Groen:2012-3}
D.~{Groen}, S.~J. {Zasada}, and P.~V. {Coveney}.
\newblock {Survey of Multiscale and Multiphysics Applications and Communities}.
\newblock {\em ArXiv e-prints arXiv:1208.6444}, August 2012.

\bibitem{Hucka:2003}
M.~Hucka, A.~Finney, H.~M. Sauro, H.~Bolouri, J.~C. Doyle, and H.~et~al.
  Kitano.
\newblock {The systems biology markup language (SBML): a medium for
  representation and exchange of biochemical network models}.
\newblock {\em Bioinformatics}, 19(4):524--531, 2003.

\bibitem{Lloyd:2004}
C.~M. Lloyd, M.~D.~B. Halstead, and P.~F. Nielsen.
\newblock {CellML: its future, present and past}.
\newblock {\em Progress in Biophysics and Molecular Biology}, 85(2–3):433 --
  450, 2004.

\bibitem{Hoekstra:2007-2}
A.~G. Hoekstra, E.~Lorenz, J.-L. Falcone, and B.~Chopard.
\newblock Toward a complex automata formalism for multiscale modeling.
\newblock {\em International Journal for Multiscale Computational Engineering},
  5(6):491--502, 2007.

\bibitem{Hoekstra:2010}
A.G. Hoekstra, A.~Caiazzo, E.~Lorenz, J.-L. Falcone, and B.~Chopard.
\newblock {\em Complex Automata: Multi-scale Modeling with Coupled Cellular
  Automata}, pages 29--57.
\newblock Understanding Complex Systems. Springer-Verlag, 2010.

\bibitem{Evans:2008}
D.~J.~W. Evans, P.~V. Lawford, J.~Gunn, D.~Walker, D.~R. Hose, R.~H. Smallwood,
  B.~Chopard, M.~Krafczyk, J.~Bernsdorf, and A.~Hoekstra.
\newblock The application of multiscale modelling to the process of development
  and prevention of stenosis in a stented coronary artery.
\newblock {\em Philosophical Transactions of the Royal Society A: Mathematical,
  Physical and Engineering Sciences}, 366(1879):3343--3360, 2008.

\bibitem{Caiazzo:2011}
A.~Caiazzo, D.~J.~W. Evans, J.~L. Falcone, J.~Hegewald, E.~Lorenz, B.~Stahl,
  D.~Wang, J.~Bernsdorf, B.~Chopard, and J.~et~al. Gunn.
\newblock {A Complex Automata approach for In-stent Restenosis: two-dimensional
  multiscale modeling and simulations}.
\newblock {\em Journal of Computational Science}, 2(1):9--17, 2011.

\bibitem{Borgdorff:2011}
J.~Borgdorff, J.-L. Falcone, E.~Lorenz, B.~Chopard, and A.~G. Hoekstra.
\newblock {A principled approach to distributed multiscale computing, from
  formalization to execution}.
\newblock In {\em Proceedings of the IEEE 7th International Conference on
  e-Science Workshops}, pages 97--104, Stockholm, Sweden, 2011. IEEE Computer
  Society Press.

\bibitem{Falcone:2010}
J.-L. Falcone, B.~Chopard, and A.~Hoekstra.
\newblock Mml: towards a multiscale modeling language.
\newblock {\em Procedia Computer Science}, 1(1):819 -- 826, 2010.

\bibitem{Zasada:2012}
S.~J. Zasada, M.~Mamonski, D.~Groen, J.~Borgdorff, I.~Saverchenko, T.~Piontek,
  K.~Kurowski, and P.~V. Coveney.
\newblock Distributed infrastructure for multiscale computing.
\newblock In {\em Proceedings of the 2012 IEEE/ACM 16th International Symposium
  on Distributed Simulation and Real Time Applications}, DS-RT '12, pages
  65--74, 2012.

\bibitem{mpwide}
D.~{Groen}, S.~{Rieder}, P.~{Grosso}, C.~{de Laat}, and P.~{Portegies Zwart}.
\newblock {A light-weight communication library for distributed computing}.
\newblock {\em Computational Science and Discovery}, 3(015002), August 2010.

\bibitem{sushi}
D.~{Groen}, S.~{Portegies Zwart}, T.~{Ishiyama}, and J.~{Makino}.
\newblock {High Performance Gravitational N-body simulations on a Planet-wide
  Distributed Supercomputer}.
\newblock {\em Computational Science and Discovery}, 4(015001), January 2011.

\bibitem{Kurowski:2010}
K.~Kurowski, T.~Piontek, P.~Kopta, M.~Mamonski, and B.~Bosak.
\newblock {Parallel large scale Simulations in the PL-Grid Environment}.
\newblock {\em Computational Methods in Science and Technology}, pages 47--56,
  2010.

\bibitem{Zasada:2009}
S.J. Zasada and P.V. Coveney.
\newblock Virtualizing access to scientific applications with the application
  hosting environment.
\newblock {\em Computer Physics Communications}, 180(12):2513 -- 2525, 2009.

\bibitem{Suter:2012}
J.~Suter, D.~Groen, L.~Kabalan, and P.~V. Coveney.
\newblock Distributed multiscale simulations of clay-polymer nanocomposites.
\newblock In {\em Materials Research Society Spring Meeting}, volume 1470, San
  Francisco, CA, April 2012. MRS Online Proceedings Library.

\bibitem{BenBelgacem:2012}
M.~Ben~Belgacem, B.~Chopard, and A.~Parmigiani.
\newblock Coupling method for building a network of irrigation canals on a
  distributed computing environment.
\newblock In Georgios Sirakoulis and Stefania Bandini, editors, {\em Cellular
  Automata}, volume 7495 of {\em Lecture Notes in Computer Science}, pages
  309--318. Springer Berlin / Heidelberg, 2012.

\bibitem{WHO_CVD_report:2012}
WHO; World Heart Federation; World~Stroke Organization, editor.
\newblock {\em {Global atlas on cardiovascular disease prevention and
  control}}.
\newblock Policies, strategies and interventions. World Health Organization,
  2012.

\bibitem{Moustapha:2001}
A.~Moustapha, A.~R. Assali, S.~Sdringola, W.~K. Vaughn, R.~D. Fish, O.~Rosales,
  G.~Schroth, Z.~Krajcer, R.~W. Smalling, and H.~V. Anderson.
\newblock {Percutaneous and surgical interventions for in-stent restenosis:
  long-term outcomes and effect of diabetes mellitus}.
\newblock {\em Journal of the American College of Cardiology},
  37(7):1877--1882, June 2001.

\bibitem{Kastrati:2000}
A.~Kastrati, D.~Hall, and A.~Sch{\"o}mig.
\newblock {Long-term outcome after coronary stenting}.
\newblock {\em Current Controlled Trials in Cardiovascular Medicine},
  1(1):48--54, 2000.

\bibitem{Jukema:2011}
J.~W. Jukema, J.~J.~W. Verschuren, T.~A.~N. Ahmed, and P.~H.~A. Quax.
\newblock {Restenosis after PCI. Part 1: pathophysiology and risk factors}.
\newblock {\em Nature Reviews Cardiology}, 9(1):53--62, September 2011.

\bibitem{Shi:2011}
Y.~Shi, P.~Lawford, R.~Hose, et~al.
\newblock Review of zero-d and 1-d models of blood flow in the cardiovascular
  system.
\newblock {\em Biomedical engineering online}, 10:33, 2011.

\bibitem{VanDeVosse:2011}
F.N. van~de Vosse and N.~Stergiopulos.
\newblock Pulse wave propagation in the arterial tree.
\newblock {\em Annual Review of Fluid Mechanics}, 43:467--499, 2011.

\bibitem{Liu:2012}
K.~Sughimoto, F.~Liang, Y.~Takahara, K.~Yamazaki, H.~Senzaki, S.~Takagi, and
  H.~Liu.
\newblock Assessment of cardiovascular function by combining clinical data with
  a computational model of the cardiovascular system.
\newblock {\em The Journal of Thoracic and Cardiovascular Surgery (in press)},
  2012.

\bibitem{Alastruey:2008}
J.~Alastruey, S.~M. Moore, K.~H. Parker, T.~David, J.~Peiró, and S.~J.
  Sherwin.
\newblock Reduced modelling of blood flow in the cerebral circulation: Coupling
  1-d, 0-d and cerebral auto-regulation models.
\newblock {\em International Journal for Numerical Methods in Fluids},
  56(8):1061--1067, 2008.

\bibitem{Mazzeo:2008}
M.~D Mazzeo and P.~V. Coveney.
\newblock {HemeLB: A high performance parallel lattice-{B}oltzmann code for
  large scale fluid flow in complex geometries}.
\newblock {\em Computer Physics Communications}, 178(12):894--914, 2008.

\bibitem{Carver:2012}
H.~B. {Carver}, R.~W. {Nash}, M.~O. {Bernabeu}, J.~{Hetherington}, D.~{Groen},
  T.~{Kr{\"u}ger}, and P.~V. {Coveney}.
\newblock {Choice of boundary condition and collision operator for
  lattice-Boltzmann simulation of moderate Reynolds number flow in complex
  domains}.
\newblock {\em Submitted to Phys Rev. E}, November 2012.

\bibitem{Botti:2010}
L.~Botti, M.~Piccinelli, B.~Ene-Iordache, A.~Remuzzi, and L.~Antiga.
\newblock An adaptive mesh refinement solver for large-scale simulation of
  biological flows.
\newblock {\em International Journal for Numerical Methods in Biomedical
  Engineering}, 26(1):86--100, 2010.

\bibitem{Groen:2012}
D.~Groen, J.~Hetherington, H.~B. Carver, R.~W. Nash, M.~O. Bernabeu, and P.~V.
  Coveney.
\newblock Analyzing and modeling the performance of the {HemeLB}
  lattice-boltzmann simulation environment.
\newblock {\em ArXiv e-prints arXiv:1209.3972}, 2012.

\bibitem{Mulder:2011}
G.~Mulder, A.~C.~B. Bogaerds, P.~Rongen, and F.~N. van~de Vosse.
\newblock The influence of contrast agent injection on physiological flow in
  the circle of willis.
\newblock {\em Medical Engineering and Physics}, 33(2):195 -- 203, 2011.

\bibitem{Carver:2012-2}
H.~B. {Carver}, D.~{Groen}, J.~{Hetherington}, R.~W. {Nash}, M.~O. {Bernabeu},
  and P.~V. {Coveney}.
\newblock {Coalesced communication: a design pattern for complex parallel
  scientific software}.
\newblock {\em ArXiv e-prints arXiv:1210.4400}, October 2012.

\bibitem{Bernabeu:2012}
M.~O. {Bernabeu}, R.~W. {Nash}, D.~{Groen}, H.~B. {Carver}, J.~{Hetherington},
  T.~{Kr{\"u}ger}, and P.~V. {Coveney}.
\newblock {Choice of blood rheology model has minor impact on computational
  assessment of shear stress mediated vascular risk}.
\newblock {\em ArXiv e-prints arXiv:1211.5292, accepted by Interface Focus},
  November 2012.

\end{thebibliography}

\end{document}